

%

\def\lunit{{\rm cm}^{-2}{\rm sec}^{-1}}
\def\YY{{\cal Y}}
\def\Ecm{E_{\rm cm}}
\def\hadrons{{\rm hadrons}}
\def\nb{{\rm nb}}
\def\ps{p_*}
\def\epem{e^+e^-}
\def\ee{\epem}
\def\Y{\Upsilon}

\def\a{\alpha}
\def\g{\gamma}

\def\sx{\sigma_x}
\def\sy{\sigma_y}
\def\sz{\sigma_z}
\def\re{r_e}
\def\Dx{D_x}
\def\Dy{D_y}

\def\L{L}

\def\avl{\left<}
\def\avr{\right>}

\def\re{r_e}
\def\g{\gamma}

\def\avl{\left<}
\def\avr{\right>}

\def\exp{\mathop{\rm exp}}

\def\Edep{E_{\rm dep}}
\def\Ptmiss{P_{\perp,{\rm miss}} }
%
\pubnum{5873}
\pubtype{T/E/A}
\date{April, 1993}
\titlepage
\title{Hadron Production in $\g\g$ Collisions as a Background
            for $\epem$ Linear Colliders\doeack}
\author{Pisin Chen, Timothy L. Barklow, and Michael E. Peskin}
\SLAC
\abstract{Drees and Godbole have proposed that, at the
        interaction point of an $\epem$ linear collider, one expects
        a high rate of
        hadron production by $\g\g$ collisions, providing
          an additional background to studies in $\epem$ annihilation.
         Using a simplified
          model of the $\g\g$ cross section with soft
            and jet-like components, we
                     estimate the expected rate of these
            hadronic events for a variety of realistic machine
                designs.}
\submit{Physical Review D}
\endpage
\chapter{Introduction}
  \REF\Beams{P.~Chen and R.~J.~Noble, SLAC-PUB-4050 (1986);
  M.~Bell and J.~S.~Bell, {\it Part.\ Accl.} {\bf 24}, 1
(1988);
R.~Blankenbecler and S.~D.~Drell,
 {\it Phys.\ Rev.\ Lett.} {\bf 61}, 2324 (1988);
P.~Chen and K.~Yokoya, {\it Phys.\ Rev.\ Lett.}
 {\bf 61}, 1101 (1988);
M.~Jacob and T.~T.~Wu, {\it Nucl.\ Phys.}
 {\bf B303}, 389 (1988);
V.~N.~Baier, V.~M.~Katkov, and V.~M.~Strakhovenko,
{\it Nucl.\ Phys.}  {\bf B328}, 387 (1989).}
\REF\CPC{
P.~Chen and V.~L.~Telnov, {\it Phys.\ Rev.\ Lett.} {\bf 63}, 1796 (1989);
R.~Blankenbecler, S.~D.~Drell, and N.~Kroll, {\it Phys.\
Rev.} {\bf D40}, 2462 (1989);
M.~Jacob and T.~T.~Wu, {\it Nucl.\ Phys.} {\bf B327},
285 (1989).}

   One of the most important issues in the design of future $\epem$
colliders is the effect of the beam-beam interaction on the
physics environment.  A linear collider operating at a center-of-mass
energy of 400 GeV and above requires a luminosity in excess of
$10^{33} \lunit$.  Such a high luminosity can only be achieved by
colliding tiny, intense bunches of electrons and positrons.  In this
circumstance, these bunches interact strongly with one another,
producing large numbers of photons and electron-positron
pairs.\refmark{\Beams,\CPC}
This effect  potentially  creates troublesome backgrounds for experiments
on $\epem$ annihilation and must be
 controlled by adjustment of the collider
parameters or the interaction region geometry.

    \REF\DG{M. Drees and R. M. Godbole, \sl Phys. Rev. Lett.  \bf 67,
               \rm 1189 (1991).}
       \REF\DGLong{M. Drees and R. M. Godbole,  DESY 92-044, BU 92/1
                          (1992).}
  In a recent set of papers, Drees and Godbole\refmark{\DG,\DGLong}
   called attention to another
potentially serious background   due to the beam-beam interaction,
in which
photons created by the bunch collision interact to produce
hadronic jets.  In some designs, the rate of this process exceeds
one jet pair per bunch crossing. Under these conditions, each $\epem$
annihilation event would be superposed on an extraneous
 system of hadronic
jets.  Thus, it is important to evaluate
this background systematically and determine its dependence on
machine parameters.

  In this paper, we will evaluate the rate of hadron and jet production
for a variety of accelerator designs which have been proposed for
500 GeV and 1 TeV $\epem$ linear colliders.  Three ingredients
are needed for such a calculation.  The first is the photon-photon
luminosity spectrum for the given linear collider   design.
The second is the
cross section for hadron production in photon-photon collisions.
The final ingredient is a realistic detector simulation to evaluate
what fraction of the produced hadrons are actually seen by the
experiments.
For  the first two of
these ingredients, we will present an explicit model
which can easily be applied to other accelerator parameters sets.
For the third, we will present some illustrative Monte Carlo
calculations.
We hope that our analysis will make it straightforward to incorporate
the constraints of the Drees-Godbole background process in any future
proposal for a linear collider.   We will also demonstrate that, with
an appropriate collider design, the Drees-Godbole  background can be
reduced to a level where it is quite unimportant.

  \REF\Ginz{I. F. Ginzburg, G. L . Kotkin, V. G. Serbo,
           and V. I. Telnov, \sl Nucl. Inst. Meth.,
           \bf 205, \rm 47 (1983);
            I. F. Ginzburg, G. L . Kotkin, S. L. Panfil, V. G. Serbo,
           and V. I. Telnov, {\it ibid.}, \bf 219, \rm 5 (1984).}
   \REF\TimGG{T. Barklow, in {\sl  Research Directions for the Decade:
        Snowmass '90}, E. L. Berger, ed.  (World Scientific, 1992).}
  \REF\Telnov{V. I. Telnov, in {\sl Physics and Experiments
 with Linear Colliders},  R. Orava, P. Eerola, and M. Nordberg, eds.
             (World Scientific, 1992).}
     \REF\UCSB{D. L. Borden, D. A. Bauer, and D. O. Caldwell,
     SLAC-PUB-5715, UCSB-HEP-92-01 (1992).}
In Section 2, we begin our study by reviewing the photon-photon
luminosity spectrum at linear colliders.  This spectrum is by
now well understood.
It  is given by  a sum of contributions
from photons radiated from electrons in the scattering process
({\it bremsstrahlung} photons)
and photons created upstream of the photon-photon collision
by the coherent action of the electric field of
one bunch on the particles of the other ({\it beamstrahlung} photons).
  The
bremsstrahlung contribution depends almost entirely
on the luminosity for
$\epem$ collisions.  The beamstrahlung contribution depends on the
accelerator parameters in a manner which is complicated but which
has by now been worked out in some detail.  We will
present an explicit parametrization
 of the  photon-photon luminosity spectrum
which incorporates these two sources in a convenient form.  We will
also review the case of a dedicated photon-photon collider, which
may be  constructed by
backscattering laser beams from the electron bunches   of an electron
linear collider.\refmark{\Ginz -\UCSB}

 \REF\ZEUS{M. Derrick, \etal\ (ZEUS Collaboration), \sl Phys. Lett.
   \bf B293, \rm 465 (1992).}
 \REF\Hone{T. Ahmed, \etal\ (H1 Collaboration), \sl Phys. Lett.
   \bf B299, \rm 374 (1993).}

The  second ingredient, the value of the
photon-photon hadronic cross section is subject to considerably
more uncertainty.  One possible model is a vector meson dominance
picture in which the photon-photon cross section  is taken to be
proportional to the $\rho$-$\rho$ cross section.  In their original
work, Drees and Godbole\refmark\DG
took a very different picture, in which
the photon-photon cross section originated from the scattering
of partons which are constituents of the two photons.  This
model leads to a cross section which is small at low energies and
increases rapidly with
energy above the center-of-mass energy of 100 GeV.
Both of these features, we believe, are unphysical.  Their use of this
model has led to considerable confusion, especially in the accelerator
physics community, as to the proper way to estimate the important
new background source to which
 they have called attention.  In Section 3, we will attempt to clarify
this issue and to present a physically reasonable scheme for estimating
the photon-photon hadronic cross sections.   As Drees and Godbole
have stressed,  two
separate questions must be addressed.  First,
what is the total cross section for hadron production?  Second, what
is the rate for hadron production accompanied by QCD jets of 5-20 GeV
transverse momentum?   Both cross sections can potentially be large
enough to lead to rates of order 1 event per bunch crossing.
In Section 3, we will argue that the total cross section is best
estimated using vector dominance ideas.  This conclusion is in accord
with recent high-energy
 measurements of the of $\gamma p$ total cross section at
 HERA.\refmark{\ZEUS,\Hone}
To estimate the cross section for events with jets, we must also
invoke the  parton-parton cross section for hard scattering.
We will evaluate this partial cross section by
introducing a very simple model, which
we call the Reference Model.  We will explain why we consider the
Reference Model a better description of the structure of jet production
than the model of Drees and Godbole.   This model follows the
essential physics of the `eikonalization' scheme of
Forshaw and Storrow.\Ref\FandS{J. R. Forshaw and J. K. Storrow
             \sl Phys. Lett. \bf B278, \rm 193 (1992).}
We have modified their model so that it contains no free parameters
and is straightforward to apply.

  In Section 4, we will combine these models of
photon spectra and the jet cross section to estimate the rates of
hadron and jet production by photon-photon reactions
for a wide variety of proposed machines.  These calculations depend on
the parameters of the machine in a quite straightforward way.
We hope that the
calculations of this section will be both sufficiently simple and
sufficiently informative
 that they can aid in the estimation
of hadronic backgrounds for future stages in the design of $\epem$
linear colliders.

 \REF\Berk{K. Berkelman, CLIC Note 164, May 1992; CLIC Note 168,
           July 1992.}
  \REF\Miyamoto{A. Miyamoto, in \sl Proceedings of the Third Workshop
   on Japan Linear Collider (JLC). \rm  A. Miyamoto, ed.   KEK
    proceedings 92-13, 1992.}
  However, as we have already noted, the full effect of hadronic
backgrounds cannot be understood without a generating hadronic
events and passing them through a realistic detector simulation.
The detectors  planned for future linear colliders typically have
holes in the forward and backward directions and substantial masking
to avoid the $\ee$ pairs produced by the beam-beam interaction.
Explicit studies of the Drees-Godbole background have shown that
much of the hadron production either is lost through these holes
or appears      at very low energy.\refmark{\Berk, \Miyamoto}
  In Section 5, we will report
at set of Monte Carlo simulations based on the model of hadron
production presented in Sections 2 and 3 of this paper.  We will
quantify the hadronic backgrounds actually detected for some
illustrative machine designs, and we will show that these backgrounds
are indeed       minor effects.

\chapter{Photon spectra from bremsstrahlung, beamstrahlung and Compton
        back-scattering}
\def\any{{\rm anything}}

 We will describe the spectra which enter $\g\g$ cross sections at
$\epem$ linear colliders in terms of a
photon-photon luminosity function $\L_{\g\g}(x_1,x_2)$.  Its parameters
 $x_1$, $x_2$ are the fractions of the total energy of the
initial electrons and positrons, respectively, carried by the
colliding photons.
 The luminosity function contributes to
cross sections as follows:
$$  \eqalign{
  \sigma(e^-( p_1) e^+ & (p_2) \to
  X + \any) \crr  &
  =  \int^1_0 dx_1 \int^1_0 dx_2
             \L_{\g\g}(x_1,x_2)
         \cdot\sigma(\gamma(x_1 p_1)\gamma(x_2p_2) \to X). \cr}
                       \eqn\Lggdef$$

  As noted in the introduction, the luminosity function receives
  contributions from two sources, beamstrahlung and bremsstrahlung,
  corresponding to  real and virtual photons.
 Assuming that the sources of the two photons are independent of
 one another, we can write the luminosity functions for an $\epem$
 collider as a sum of components:
 $$   \L_{\g\g}(x_1,x_2) =   f_v(x_1) f_v(x_2)  + \bigl[
  f_v(x_1) f_r(x_2) + f_r(x_1)
  f_v(x_2) \bigr] + f_r(x_1) f_r(x_2) .
                \eqn\lumismade$$
In this equation, $f_v(x)$ is a modification of the
Weizs\"acker-Williams
distribution for radiation in a collision process, $f_r(x)$ is the
average of the beamstrahlung spectrum  over the process
of interpenetration of the $e^-$ and $e^+$ bunches.  In the cross term,
there may be a geometrical suppression  of the virtual photon
distribution.  This effect is important in $\epem$ pair creation
at the interaction point in
linear colliders.\Ref\georeduc{
    A. E. Blinov, A. E. Bondar, Yu. I. Eidelman, et al.,
    {\sl Phys. Lett.} {\bf B113}, 423 (1982);
    Yu. A. Tikhonov, {\it Candidates's Dissertation \rm},
    Inst. Nucl. Phys., Novosibirsk (1982);
    V. N. Baier, V. M. Katkov and V. M. Strakhovenko,
      {\sl Sov. J. Nucl. Phys. \bf 36 \rm}, 95 (1982);
    A. I. Burov and Ya. S. Derbenev,  INP Preprint 82-07,
    Novosibirsk (1982);
    G. L. Kotkin, S. I. Polityko and V. G. Serbo,
      {\sl Sov. J. Nucl. Phys. \bf 42 \rm}, 440 (1985);
   P. Chen, T. Tauchi and D. V. Schroeder,
      ``Pair Creation at Large Inherent Angles'',
      in {\sl Research Directions for the Decade $\cdot$ Snowmass 1990},
      World Scientific (1992).}
  However, the same logic       predicts
that this effect is negligible for the process considered here.

   To compute the jet production cross section at a jet transverse
momentum of order $Q$, Drees and Godbole have argued that one should
use a modified version of the standard Weizs\"acker-Williams formula.
The standard formula integrates over all photon transverse momenta.
However, only those photons which are off-shell by less than $Q^2$,
can produce jets with transverse momentum of order $Q$  with an
unsuppressed rate.  In addition, only a fraction $c_v$ of the partons
in these photons will be off-shell by an amount less than $Q^2$.
By integrating the formula for the equivalent photon distribution
given by Brodsky, Kinoshita, and
Terazawa\Ref\BKT{S. Brodsky, T. Kinoshita, and H. Terazawa, \sl
Phys. Rev. \bf D4, \rm 1532 (1971).}  up to $Q^2$ and applying the
additional suppression factor, we obtain
$$ \eqalign{
  f_v(x,Q,E) = c_v  \cdot  {\alpha\over 2 \pi x}&
          \bigl[ (1 + (1-x)^2)        (\log{Q^2\over m_e^2}-1 )\cr
   & + {x^2\over 2} \bigl(\log{(1-x)\over x^2} + 2 \bigr) +
      {(2-x)^2\over 2} \log{ (1-x)\over (Q^2/E^2 + x^2)}  \bigr].\cr}
      \eqn\fvdef$$
where $E$ is the electron beam energy and\refmark\DG
$$    c_v = 0.85 .             \eqn\cval$$
The distribution \fvdef\ modifies a simple dependence  proportional
to $\log Q^2$ to include the correct enhancement at small $x$
and suppression at large $x$ from the electron kinematics.

\REF\Noble{R.~J.~Noble, {\sl Nucl. Instr. Meth.} {\bf A256}, 427 (1987).}
\REF\Chen{
P.~Chen, ``An Introduction to Beamstrahlung and Disruption'', in
{\it Frontiers of Particle Beams}, Lecture Notes in Physics {\bf 296},
Springer-Verlag, 1988.}
\REF\Holle{R.~Hollebeek, {\sl Nucl. Instr. Meth.} {bf 184}, 333 (1981).}
\REF\CCY{P.~Chen and K.~Yokoya, {\sl Phys.\ Rev.}  {\bf D38},
987 (1988); K.~Yokoya and P.~Chen,
    ``Beam-Beam Phenomena in Linear Colliders'',
in {\it Frontiers of Particle Beams: Intensity Limitations}, Lecture
  Notes in Physics {\bf 400}, Springer-Verlag, 1992.}
\REF\Yok{K.~Yokoya, KEK Report 85--9 (1985).}
\REF\GG{P.~Chen, ``Beamstrahlung and the QED, QCD Backgrounds in Linear
      Colliders'', SLAC-PUB-5914 (1992); submitted to the Proceedings
      of the 9th International Workshop on Photon-Photon Collisions
      (Photon-Photon '92).}
\REF\Pisinonethird{J. B. Rosenzweig and P. Chen, in \sl Proc. of IEEE
  Particle Accelerator Conference 91CH3038-7, \rm M. Allen, ed.
      (IEEE Service Center, 1991).}
\REF\Pisin{
P.~Chen, {\sl Phys. Rev.} {\bf D46}, 1186 (1992).}

In contrast to bremsstrahlung, beamstrahlung occurs in the situation
where the scattering amplitudes between the radiating particle and the
target particles within the characteristic length add coherently.
Typically for the beam-beam collision in linear colliders there can be
 over a million target particles involved within the coherence
length. The process
can therefore be well described in a semi-classical calculation
where the target particles are replaced by their collective EM fields.

 High energy $\epem$ beams generally follow Gaussian
distributions in the three spatial dimensions, and their local field
strength varies inside the beam volume.
In the weak disruption limit, where particle motions have small
deviations from the $z$ direction,
it is possible to
integrate the radiation process over this volume and derive relations
which depend only on averaged, global beam parameters.
The overall beamstrahlung intensity is controlled by
a global {\it beamstrahlung parameter\/},\refmark{\Chen,\Noble}
$$
   \Y_0 = \g{\avl B\avr \over B_c} =
    {5 \over 6} {\re^2\g N \over
      \a\sz(\sx+\sy)}\quad,
        \eqn\upsilnom
$$
where $\avl B \avr$ is the mean
 electromagnatic field strength of the beam,
 $B_c=m_e^2/e\simeq 4.4\times 10^{13}$ Gauss is the Schwinger critical
field, $N$ is the total number of particles in a bunch,
$\sx, \sy, \sz$ are the nominal sizes of the Gaussian beam,
$\g$ is the Lorentz factor of the beam,
$r_e$ is the classical electron radius,
and $\a$  is  the fine structure constant.
The collective fields in the beam also deform the other beam during
collision, by an amount controlled by
global {\it disruption parameters\/},  which may be different in the two
transverse directions:\refmark{\Holle,\CCY}
$$
  D_{x,y}={2N r_e\sz \over \g\sigma_{x,y}(\sx+\sy)}\quad.
            \eqn\disrup
$$

 In the most general designs for linear colliders, the photon spectrum
due to beamstrahlung is not a factorized function of the electron
and positron sources and depends on the detailed evolution of the
bunches in the collision process.  In general, then, the spectrum of
radiation must be computed by detailed simulation.\refmark{\Yok,\Noble}
However, typical beams in linear colliders are very long and narrow.
Since all particles oscillate within the focusing potential that is
defined by the geometry of the oncoming beam, the oscillation amplitudes
are small compared with their periodicity in $z$. Then the
 assumption of small deviations from the $z$ direction
remains approximately valid. The
main effect of disruption on beamstrahlung is the change of effective
EM fields in the bunch due to the deformation of the transverse beam
sizes. Thus, beamstrahlung is in practice still factorizable even under
a non-negligible disruption effect, if one computes its magnitude
using an effective beam size which takes the global disruption into
account.

The proper value of this
effective beam size can be found from the
luminosity enhancement factor, defined as the ratio of the effective
luminosity to the nominal luminosity due to the change of beam size:
$$
  H_{_D}\equiv {\bar{\cal L} \over {\cal L}}={\sx\sy \over
   \bar{\sigma}_x\bar{\sigma}_y}\quad,
              \eqn\lumenhan
$$
The luminosity enhancement factor is calculable analytically only in
the $D\ll 1$ limit.
 Beyond this limit the dynamics of beam-beam
interaction becomes nonlinear, and one must use simulations.
 From the results of these simulations, we can extract
 scaling laws for $H_{_D}$  and thus for the effective beam size.
For the case of round beams ($\sx/\sy = 1$), simulations produce
the behavior:\refmark\CCY
$$
   H_{_D}=1+D^{1/4}\Bigl({D^3 \over 1+D^3}\Bigr)
          \Bigl\{\ln(\sqrt{D}+1)+2\ln(0.8/A)\Bigr\}\quad,
             \eqn\hdscaling
$$
where $A=\sz/\beta^*$, and $\beta^*$ is the Courant-Snyder
 $\beta$-function at the interaction point. This scaling
law is valid to about 10\% accuracy. Thus for round beams,
 the effective beam size is roughly given by
$\bar{\sigma} = \sigma H_{_{_D}}^{-1/2}$.

In realistic designs for high-energy $\ee$ colliders, the beams
are intentionally made quite flat, with $R= \sx/\sy$ greater than
5 and as large as 100 in some designs.  In this case, there are
separate $\beta^*$ values and separate disruption parameters  \disrup\
 for the $x$ and $y$ directions.  Typically, $H_{Dx}$, computed from
 \hdscaling \ with $D = D_x$ and $A = \sz/\beta_x^*$, is close to 1,
 while $H_{Dy}$, computed from \hdscaling \ using the $D_y$ and
$\sz/\beta^*_y$, is large.  Since the field strength in a flat
charge distribution is mainly determined by $\sx$, this means that the
disruption effect and its enhancement of beamstrahlung will be
relatively mild. However, it turns out that the effect of $\sy$ is
quantitatively important and cannot be neglected.

   We therefore suggest the following prescription for computing the
effective beamstrahlung parameter:  Let\refmark\GG
$$
  \bar{\sigma}_x =   \sx H_{_{D_x}}^{-1/2}\quad,
  \qquad     \bar{\sigma}_y =    \sy H_{_{D_y}}^{-1/3}.
    \eqn\sigmaeff
$$
The exponent $(1/3)$ in the second term is determined from computer
simulations for very flat beams
 in which the horizontal particle motion is ignored
(ref. \CCY);  a theoretical basis for this scaling law has been
proposed in ref. \Pisinonethird.
 Then the effective beamstrahlung
parameter is given by
$$
   \Y= {5 \over 6} {\re^2\g N \over
      \a\sz(\bar{\sigma}_x+\bar{\sigma}_y)}\quad.
        \eqn\upsiloneff
$$
This prescription gives beamstrahlung spectra which agree with the
simulation results to an accuracy of 10\% for colliders with
flat beams ($R > 5$).

  Once one has an effective value of
   the beamstrahlung parameter, it is straightforward to derive
   the photon spectrum.\refmark\Pisin
The number of soft photons radiated
per unit time, calculated by the classical theory of radiation, is
$$
       \nu_{cl}={5 \over 2\sqrt{3}}{\a^2 \over \re \g}\Y\quad.
        \eqn\nucl
$$
Note that for a given field strength $\nu_{cl}$ is independent of the
particle energy. This expression applies to the infrared limit of the
spectrum where photon energies approach zero.   For a hard photon,
up to the initial energy of the electron, the quantum mechanical
calculation gives a more general formula:
$$
        \nu_{\g} =   \nu_{cl}[1+\Y^{2/3}]^{-1/2}\quad.
         \eqn\nug
$$
In a multi-photon radiation process,
 it was found useful to
 introduce a linear interpolation between these two values.  Let
$x$ be the energy fraction of the initial electron carried by the
photon. Then define
$$
   \bar{\nu}(x)
   ={1 \over 1-x}\int_x^1dx'[x'\nu_{cl}+(1-x')\nu_{\g}]
                            ={1 \over 2}\Big[(1+x)\nu_{cl}
                            +(1-x)\nu_{\g}\Big]\quad.
           \eqn\nubar
$$
With these basic parameters introduced, $f_r(x)$ is given by\refmark
{\Pisin}
$$
          f_r(x)={1 \over \Gamma(1/3)}\Bigl({2 \over 3\Y}\Bigr)^{1/3}
                 x^{-2/3}(1-x)^{-1/3}\exp\Big[-{2x\over
                3\Y(1-x)}\Big] \cdot G(x)\ ,
       \eqn\frval
$$
where  $\Y$ is given by \upsiloneff,
$$ \eqalign {
        G(x)&={1-w \over g(x)}\Big\{1-{1 \over
             g(x)n_{\g}}\Big[1-e^{-g(x)n_{\g}}\Big]\Big\}
             +w\Big\{1-{1 \over
             n_{\g}}\Big[1-e^{-n_{\g}}\Big]\Big\} \quad , \quad
        \cr \noalign {\vskip 10pt}
        g(x)  &=  1-{\bar{\nu} \over \nu_{\g}}(1-x)^{2/3}\quad, \cr}
 \eqn\gparam
$$
and
$$
        w= {1 \over 6}\sqrt{3\Y\over 2} \quad ,  \quad\quad
        n_{\g}=\sqrt{3}\sz\nu_{\g}.
       \eqn\moredef
$$
$n_{\g}$ is the mean number of photons radiated per electron throughout
the collision. The approximations are valid for $\Y \lsim 5$.

 So far, we have been discussing the photon spectra associated with
 linear colliders operating in a mode to study $\epem$ collisions.
 It is also possible to run a linear collider in a mode dedicated
to the study of $\g\g$ collisions, by backscattering a laser beam
from each electron beam just before the collision point. The luminosity
for photon-photon collisions should be essentially equal to the
design luminosity for $\epem$ collisions, without the enhancement
factor \lumenhan.
Ten years ago, Ginzburg, Kotkin, Serbo, and Telnov\refmark{\Ginz}
studied this possibility in some detail and displayed many interesting
characteristics of the photon-photon collider.  In particular, they
computed the luminosity spectrum   of each photon beam. Ignoring
polarization  effects,
  $$    f_c(x) = {1\over {\cal N}} \bigl[ 1 - x + {1\over 1- x}
         - {4x \over X(1-x)} + {4 x^2 \over X^2 (1-x)^2} \bigr] ,
   \eqn\backscatter$$
where
$$  {\cal N}  = \bigl( 1- {4\over X}  - {8\over X^2}\bigr)\log(1+X)
          + {1\over 2} + {8\over X} - {1\over 2(1+X)^2}   .
          \eqn\thenorm$$
The parameter
  $X$ is related to the center of mass energy of the electron-laser
photon collision:
$X = (E_{\rm cm}/m_e)^2$, and $x$ is restricted to $x < X/(1+X)$.
   Telnov\Ref\Telnov{V. I. Telnov, \sl Nucl. Inst. Meth. \bf A294, \rm
                      72 (1990).}
     has argued that the
optimal value of $X$
is $X  = 2 + \sqrt{8} \approx 4.83$, and we will use this value here.
The luminosity function for the $\g\g$ collider is then simply
$$
   L_{\g\g}(x_1,x_2)= f_c(x_1)f_c(x_2).
    \eqn\gglumspec
$$

   The formulae tabulated in this section give
 a complete and rather straightforward method for computing the
photon spectrum relevant to background processes at future linear
colliders.

\chapter{The $\g\g$ total cross section}

  In order to compute hadronic backgrounds due to the photon spectrum
described in Section 2, we must fold this spectrum with a reasonable
theory of the photon-photon hadronic cross section.  Unfortunately,
this cross section has been measured only at very low energies---energies
below 20 GeV in the center of mass.  The extrapolation of these
measurements even to 100 GeV in the center of mass depends on the
theoretical models.  In this section, we will describe a simple, specific
model which we propose should be taken as a reference.

\REF\Amaldi{U. Amaldi, \etal, \sl Phys. Lett. \bf
         66B, \rm 390   (1977).}
   \REF\PLUTO{Ch. Berger, \etal\ (PLUTO collaboration), \sl Phys. Lett.
      \bf 149B, \rm 421 (1984).}
         \REF\TPC{H. Aihara, \etal\ (TPC/Two-Gamma collaboration),
               \sl Phys. Rev. \bf D41, \rm 2667 (1990).}
      \REF\NOVO{S. E. Baru, \etal\ (MD-1 collaboration),  Novosibirsk
            preprint IYF-91-52 (1991).}
  The simplest model of the energy-dependence of the photon-photon
hadronic cross section is that given by vector meson dominance.  In this
model, the photon is considered to resonate, with some amplitude, to a
hadronic state such as the $\rho$.  Then the photon-photon total
cross section should be proportional to the $\rho$--$\rho$ total cross
section as a function of energy.  In practice, among the hadronic total
cross sections, only the $pp$ and
$p\bar p$ cross sections are measured  above 30 GeV in the center of
mass.  We will estimate the energy-dependence of the photon-photon
total cross section by averaging these to remove the effects of
baryon exchange.  Using the parametrization of
Amaldi, \etal\refmark\Amaldi
  (which continues to fit
the more recent SPS and Fermilab data), we have
$$  \sigma(\g\g \to {\rm hadrons}) = \sigma_0 \cdot
\Bigl\{ 1 + (6.30 \times 10^{-3}) [\log(s)]^{2.1} + ( 1.96)
 s^{-0.37} \Bigr\} ,
   \eqn\Amaldifit$$
   where $s$ is given in (GeV)$^2$.
The same formula, with $\sigma_0 = 80\ \mu$b, describes the new
high-energy determinations of the $\gamma p$ total cross section
from HERA.\refmark{\ZEUS,\Hone}  To describe $\g\g$ scattering,
the constant may be adjusted so that $\sigma(\g\g) =
[\sigma(\g p)]^2/\sigma(pp)$ in the region of approximately constant
cross sections at $\Ecm \sim 30$ GeV:
$$   \sigma_0 =  200\ \nb                 \eqn\sigzeroval$$
     \FIG\Amalf{The parametrization of the photon-photon total
             hadronic cross section, eq. \Amaldifit, compared to
             data from refs. \PLUTO (circles), \TPC
   (squares), \NOVO (dots).}
The formula \Amaldifit\ is plotted in Figure \Amalf\ and compared to
direct determinations of the $\g\g$ hadronic cross
section.\refmark{\PLUTO,\TPC,\NOVO}
Comparing $\sigma(\g p)$ to $\sigma(\pi p)$, we conclude that the
photon is a hadron a fraction (1/400) of the time.

  A second model of the photon-photon cross section is one based
on parton-parton scattering.
    Many authors have speculated that the hard QCD
processes can make a significant contribution
to the total cross section in hadron-hadron scattering at high energies.
Drees and Halzen\Ref\Halzen{M. Drees and F.
Halzen, \sl Phys. Rev. Lett. \bf 61, \rm 275 (1988).}
 proposed that parton-parton scattering could be the dominant
process in the photon-hadron cross section  above 100 GeV in the center
of mass, and that this mechanism would lead to photon-hadron cross
sections which rise much faster than  \Amaldifit.  This theory of the
photon-hadron cross section was then taken over by Drees and Godbole to
describe the photon-photon hadronic cross section.  However, this
theory has been criticized in both contexts by many authors.  Let us
first write out a simple, quantitative version of the
Halzen-Drees-Godbole theory, and then explain how this theory should
be used in the calculation of  hadronic backgrounds.

   At the most naive level, the cross section for hadron production by
hard parton-parton scattering is given by folding the parton scattering
cross sections computed in QCD with the experimentally determined
parton distributions.  In general, this cross section is infrared
divergent and requires a cutoff at low momentum transfer or transverse
momentum.  Halzen noted that one can obtain cross sections of the order
of the expected total cross section for hadron production if one takes
this cutoff to be a few GeV.  In this calculation, the most important
effect comes from gluon-gluon scattering at small momentum fractions.

Let us
define the jet yield $\YY(\ps)$ as the expected number of
jets  with $p_\perp > \ps$, divided by the luminosity.
The simplest hard-scattering theory of the total cross section would be
to take
$$     \sigma(\g\g)
 =  \sigma_0  +   {1\over 2}  \YY(\ps) ,   \eqn\naive$$
where $\sigma_0$ is a constant soft-scattering cross section and
the cutoff $\ps$ is taken sufficiently large that events contributing
to the jet yield are not  also accounted as part of $\sigma_0$.  Let us
first describe how we evaluate  $\YY(\ps)$, and then discuss its
relation to the total cross section.

 We compute
 $\YY(\ps)$ from the formula:
$$
  \YY(\ps) =   \int^1_0 dz_1  F(z_1)  \int^1_0
 dz_2  F(z_2) \int^1_{-1}   d\cos\theta
        {d\sigma\over d\cos\theta}(gg\to gg) \cdot \theta(p_\perp-\ps).
             \eqn\Yieldcalc$$
  In this formula, $\theta$ is the center-of-mass parton-parton
   scattering angle. We take the parton distribution $F(z)$
to be the sum of gluon and quark distributions
$$    F(z)   = f_g(z) + {4\over 9} \sum_i [f_{qi}(z) + f_{\bar q i}(z)]
  \eqn\myglue$$
with the appropriate coefficient that we can
 approximate all of the parton cross sections by the gluon-gluon
cross section:
$$
  {d\sigma\over d\cos\theta}(gg\to gg) =  {9\over 16}{\pi \alpha_s^2
     \over \hat s} \Bigl[{( 2 + \cos^2\theta)^3\over \sin^4\theta}
        \Bigr] ,
\eqn\leadinggg$$
where $\hat s  = z_1 z_2 s$ is the square of the gluon-gluon center
of mass energy.  The coupling constant
 $\alpha_s$ is evaluated at the momentum scale $p_\perp$.  We compute
 $\alpha_s$ from leading-order evolution with 4 flavors and $\Lambda
 = 400$ MeV ($\alpha_s$(3 GeV) = 0.37), the convention of Drees and
Godbole.

  For the parton distributions of the photon, we use the parametrization
  of Drees and Grassie.\Ref\DGrass{M. Drees
   and K. Grassie, \sl Z. Phys. \bf C28, \rm 451 (1985).}
  The gluon distribution in the photon is
   poorly known experimentally. However, this distribution should be
calculable theoretically to an accuracy of about 30\%\  by integrating
the Altarelli-Parisi equations, taking as an initial condition the
gluon distribution in a meson, multiplied by the probability (in vector
dominance) that the photon resonates with a meson.  Although Drees and
Grassie took their initial condition from the early photon-photon
scattering data from \hbox{PETRA},
their result actually agrees quite well
with the result of this more theoretical criterion.

 The QCD result
for the gluon-gluon scattering cross section at low momentum transfer
has much larger uncertainties.   First of all, the lowest order QCD
result receives large perturbative corrections.  There are further
corrections which come from outside the standard leading-log
diagrams of QCD. On the one hand, the summation of diagrams relevant
to multiple   gluon production reveals that gluon-gluon scattering
is controlled by a Regge pole which increases the cross section
proportional to a (small) power of the gluon-gluon center-of-mass
energy.\Ref\Lipatov{E. A. Kuraev, J. N. Lipatov, and V. S. Fadin,
  \sl  Sov. Phys. JETP    \bf 45, \rm 199 (1977);  Ya. Ya. Balitsky
   and L. N. Lipatov, \sl Sov. J. Nucl. Phys. \bf 28, \rm 822 (1978).}
  On the other hand, because
the photon is a total color singlet, the amplitudes for creating
 low transverse momentum gluons should exhibit cancellations
 between the various color sources.\Ref\Collins{A second possible
 source of
    suppression, the saturation of the gluon evolution at small
     $z$, has been discussed by  Collins and Ladinsky, ref. \CL.
   However, this effect is small in the energy region we consider in this
              paper.}
  Both classes of corrections
are beyond the scope of this paper.
{}From here on, we will consider \Yieldcalc\ as a standard reference
point for the calculation of jet production.  We expect that it yields
a calculation of $\YY(\ps)$ up to an uncertainty of about a factor of 2.

  \FIG\Yields{Jet yields predicted by the formula \Yieldcalc, for
          $\ps = 1.6$, 3.2, 5 and 8 GeV, shown as a function of the
            $\g\g$ center of mass energy.  The dotted curves show the
                 parametrization of this quantity given in \Psparam.}
   This said, we present in Fig. \Yields\ the result of evaluating
$\YY(\ps)$.  From the simplest point of view, this is a theory of the
photon-photon hadronic cross section:  $\sigma \sim (1/2) \YY(\ps)$,
for an appropriately chosen value of $\ps$.  This is, in fact, the
theory applied by Drees and Godbole, with the parameter choice
  $\ps = 1.6$ GeV.\refmark\DGLong   Notice
that
for any value of $\ps$, this prediction for the cross
 section rises much faster at
high energy than the expectation from \Amaldifit.
  In addition,
this prediction for the cross section is very small at low energy, since
it does not include the effects of soft hadronic reactions.
  The dependence
of the  jet yield on energy and  $\ps$ is well described by the
parametrization
$$   \YY(\ps,\Ecm)  =  A_1{ (\Ecm)^{A_2} \over (A_3 + \ps)^2}
    \exp\Bigl\{ -  { B(\ps)\over (\Ecm - \ps)^{C(\ps)}}  \Bigr\} ,
              \eqn\Psparam$$
with $A_1 = 4000$, $A_2 = 0.82$, $A_3 = 3.0$, and
$$   B(\ps) = 14.2 \tanh(0.43 \ps^{1.1})\ ,\quad\quad C(\ps) = 0.48/
           \ps^{0.45}\ .
     \eqn\Psconsts$$
This parametrization fits our numerical evaluation
 to  within 20\% accuracy for $\ps < 10$ GeV
and $\Ecm < 10$ TeV.   As we have emphasized, the numerical evaluation
itself is considerably more uncertain.   We used
this parametrization
in the computations reported in Section 4.

\REF\Durand{L. Durand and
H. Pi, \sl Phys. Rev. Lett. \bf 58, \rm 303 (1987),
    \sl Phys. Rev. \bf D40, \rm 1436 (1989).}
    \REF\CL{J. Collins and G. Ladinsky, \sl Phys.
         Rev. \bf D43, \rm 2847 (1991).}
  \REF\Alekh{S. I. Alekhin, \etal, CERN-HERA 87-01 (1987).}
  However, it has been argued that the photon cross sections
cannot rise as fast as the jet yield is predicted to rise in Fig.
\Yields.\refmark{\Durand, \CL}
  The easiest  way to argue to this conclusion is to redo the analysis
just described
              for $p\bar p$ collisions and compare the results to
    \FIG\Comparepp{(a)
    Comparison of the jet  yield in $p\bar p$
 collisions to the observed total cross section.  The data is taken from
   ref. \UAone.  The upper set of data points represents measurements
of the inelastic $p\bar p$ cross section; these measurements are
well fit by the formula of ref. \Amaldi.  The lower set of data points
represents the UA1 measurements of the jet cross section, as described
in the text.  The two curves show the energy-dependence of
(1/2)$\YY(\ps)$
      for $p\bar p$ collisions, for $\ps = 1.6$, 3.2 GeV.
(b)    Comparison of the jet  yield in $\gamma p$
 collisions to the observed total cross section.  The data is taken from
   refs. \Alekh, \ZEUS, \Hone.  The smooth curve through these points
   is proportional to    \Amaldifit.
  The two rising curves show the energy-dependence of
(1/2)$\YY(\ps)$
      for $\gamma p$ collisions, for $\ps = 1.6$, 3.2 GeV.}
the data on the $p \bar p$ total cross section.
                This comparison is shown in Fig. \Comparepp(a).  Notice
  that the jet yield calculation using the Drees-Godbole value of
       $\ps$ is completely incompatible with the $p \bar p$ total
  cross section in a region where this cross section is well measured.
A  similar comparison  can now be made in photoproduction following the
new HERA measurements, and this in shown in Fig. \Comparepp(b).

   In addition to the total cross section,
  The UA1 experiment has reported measurements of the cross section
for events with jet activity, by counting events with a fixed deposition
of transverse energy in a circle of radius 1 in the plane of
rapidity and azimuthal angle.\Ref\UAone{C. Albajar, \etal\ (UA1
 experiment), \sl Nucl. Phys. \rm B309, \rm 405 (1988).}
In this paper, the experimenters argued that such minijet events are
well defined only for values of the transverse energy of a cluster
above 5 GeV.  In Fig. \Comparepp(a), we show their results for the
cross section for producing clusters of 5 GeV transverse energy, and
the comparision of this cross section to half the jet yield for a
parton transverse  momentum cutoff $\ps = 3.2$ GeV.  At the time of these
mesurements, Pancheri and Srivastava\Ref\PVS{G. Pancheri
   and Y. Srivastava, \sl Phys. Lett. \bf B182, \rm 199 (1986).}
 pointed out that this
cross section could be fit by a simple QCD estimate with a value of
$\ps$ reduced from the observed transverse energy.  The comparision
shown in Fig. \Comparepp\ fixes the size of this reduction for the
Drees-Godbole conventions.  To estimate the cross section for events
with clusters of 10 GeV transverse energy, we will use $\ps = 8$ GeV.

The idea that minijets with only 5 GeV of transverse energy are
produced independently of the underlying minimum-bias multiple particle
production is still controversial.  It is possible that a model with
incoherently produced jets makes sense at
values of the transverse energy of 10 GeV or above.  When we evaluate
jet cross sections later is this paper, we will also illustrate the
dependence of our results on the transverse momentum cutoff $\ps$.

  We will now argue that, in photon-photon collisions, we should see
the same disagreement between the actual total cross section and the
jet yield calculation at high energy.
At low energies, photon-photon collisions have
 an approximately constant hadronic cross section
from vector dominance: each photon resonates, with a certain probability
to a hadron, and these hadrons collide with a certain total cross
section.  Taking the probability that the photon is a hadron to be the
value (1/400) given above, and taking the maximum hadronic cross section
to be that of a disc of radius 1 fm, we obtain  an estimate
$$    \sigma_T(\g\g \to \hadrons) \sim  300\  \nb
      \eqn\theest$$
which is in reasonable agreement with \Amaldifit.   In order to produce
a significantly larger cross section, either the photon must become
larger or it must become a hadron with higher probability.  Resolving
the hadronic components of the photon into partons does not increase
the size of the photon.  Altarelli-Parisi evolution can create new
hadronic components of the photon, through the diagram in which the
photon off shell by an amount $Q$ splits to a $q\bar q$ pair. This
diagram has a substantial effect on the total number of gluons in the
photon, but it has only a small effect on the photon's hadronic
cross section, since the new hadronic component has the very small
size $\pi/Q^2$.   It is possible to explain a slowly rising
cross section by making a model in which the soft hadron is a grey
scattering distribution which becomes black as the gluon-gluon scattering
becomes important.  As the disk becomes black, the effect of
gluon-gluon scattering on the total cross section must turn off.
This physical effect can be implemented in a calculational scheme
called `eikonalization'.
  For the case of $\g p$ scattering, models of
this sort have been constructed by Durand and Pi,\refmark\Durand
Forshaw and Storrow,\Ref\FSph{J. R. Forshaw and J. K. Storrow,
   \sl Phys. Lett. \bf B268, \rm  116 (1991), (E) \bf B276, \rm 565
      (1992).}  and Fletcher,
      Gaisser and Halzen.\Ref\GHz{R. S. Fletcher, T. K. Gaisser, and
    F. Halzen, \sl Phys. Rev. \bf D45, \rm 377 (1992).}
  Forshaw and Storrow have also
written an eikonalized model of the $\g\g$ cross section.\refmark\FandS
Qualitatively,
 these eikonalized models have a slowly rising total cross section
     \REF\HB{M. M. Block, R. Fletcher
     F. Halzen, B. Margolis, and P. Valin, \sl Phys. Rev. \bf D41,
      \rm 978  (1990).}
similar to that of \Amaldifit. An example of such a model which
fits the rise of the $pp$ cross section has been given in ref. \HB.
  On the other hand, it is possible that parton hard scattering
has nothing to do with the observed rise in the $pp$ cross section
at high energy.  In this paper, we will adopt the most straightforward
course, that of taking
the formula \Amaldifit \ literally as a first approximation to the
energy-dependence of the cross section for hadron production in $\g\g$
collsions.

    However, we are also interested to know the cross section for
hadronic reactions which contain hard QCD jets.   It is
quite possible that ordinary, low-$p_\perp$ hadronic events produce
little complication when superposed on high-energy $\epem$ annihilation
events, but that hadronic events with jets produce troublesome
complications.  Thus, we need to estimate
 backgrounds    from events with jet production.
  We emphasize
that we are concentrating on the case of jets with transverse momentum
below 20 GeV which appear as the result of a second collision at the
same beam crossing as the $\epem$ annihilation.   Above this transverse
momentum, parton-parton scattering decreases in importances as a source
of hadronic jets relative to quark-photon and direct photon-photon
scattering processes (the processes Drees and Godbole call
`once-resolved' and `direct').\refmark\DG
  However, these latter events are too
rare to appear superposed on a signficant number of $\epem$ annihilation
 events.

   To a first approximation, the jet yield $\YY(\ps)$ computed from
\Yieldcalc\ should be a valid estimate of the total number of jets
produced even when  the jet yield substantially overestimates the total
hadronic cross section.  The reason for this is that the individual
parton-parton interactions are relatively weak, and it is only because
there are many gluons in a hadron that the sum of these cross sections
saturates the geometrical limit on the cross section.   In other words,
those events in which the hadronic disks overlap typically contain
a soft interaction plus gluon-gluon scatterings;  if $\YY(\ps)
\gg \sigma$,
typical encounters contain many individual gluon-gluon collisions.
If we assume that these collisions are completely independent, we
would expect the number of pairs
of jets per event to  follow a Poisson
distribution, such that the mean number of jets per event is
$$    \VEV{n_{\rm jet} } =    \YY(\ps)/\sigma .
   \eqn\meanjet$$
The cross section for events with jets of $p_\perp > \ps$, in this
model, is
$$     \sigma(\ps)  =  \sigma \cdot \bigl\{1 - \exp\big[-\YY(\ps)/2\sigma
                      \big] \bigr\}.
   \eqn\sigmaprob$$
If the mechanism of scattering changes as a function of the impact
parameter, as is true in eikonal models, there will be small corrections
to this simple model.  We will ignore them.

   The combination of these ideas has an interesting implication.
$\YY(p)$ increases much more rapidly with energy than $\sigma$.
However, in this picture, the main effect of the increase in $\YY(\ps)$
is not to
increase the  hadronic   cross section but rather to increase the
number of jets per event.  For photon-photon collisions, and for
   \FIG\TimeStruct{Time structure of $\ee$ reactions in a linear
   collider.  The dots represent individual bunch crossings.
     In the naive model (a), the minijets are distributed evenly
          among bunch crossings.  The model (b) has a much smaller
             $\g\g$ hadronic cross sections, but the same large
               value of the jet yield.}
hadron-hadron collisions, above 1 TeV in the center of mass, we
expect that the typical event is bristling with jets of 10 GeV
transverse momentum.   In Fig. \TimeStruct, we illustrate the time
structure of events at an $\epem$ collider in a naive model
and in what we feel is a more correct model of jet production.
The latter case casts the problem of hadronic jets underlying $\epem$
annihilation events in a quite different form, and one which is
   \FIG\Njets{Number of  jets with transverse momentum greater than
     $\ps$
   per hadronic $\g\g$ event,  for
$ \ps  = 1.6$, 3.2, 5, 8 GeV, according to the model of eq. \meanjet.
    The ordinate is the $\g\g$ center of mass energy.}
probably much easier to ameliorate.
 In Fig. \Njets, we show the energy dependence
of the mean number of jets in $\g\g$ collisions with hadron
production, according to our model, for various values of the
   \FIG\NCross{Cross sections for hadron production in $\g\g$
collisions accompanied by jets of transverse momentum greater
than $\ps$, for
$ \ps  = 1.6$, 3.2, 5, 8 GeV, according to the model of eq. \sigmaprob.
    The ordinate is the $\g\g$ center of mass energy.}
transverse momentum.   In Fig. \NCross, we show the corresponding
predictions for the total cross section for a $\g\g$  collision
to produce events with parton scattering at these values of
transverse momentum.  We have already noted that the results for
the lowest value of $\ps$ is probably academic, since such small
minijets cannot be distinguished in hadron-hadron collisions.
The two curves with $\ps = 3.2$, 8 GeV
correspond to events with clusters of 5, 10 GeV transverse
energy.

Since Fig. \Njets\ predicts a relatively large number of jets per
hadronic event, one might hope that multiple jet events could be
recognized experimentally in $\gamma p$ or $p\bar p$ collisions at
accessible energies.  Unfortunately, our model gives fewer
jet-like events in these processes, since the gluon distribution in
the proton is softer than that in the photon.  For $\ps = 3.2$ GeV,
we estimate an average of 0.15  jet pairs for $\gamma p$ collisions at
200 GeV, and an average of 0.6 jet pairs for $p\bar p$ collisions at
2 TeV.  However, we expect  2 jet pairs per minimum bias event at
the SSC energy of 40 TeV, so that the phenomenon of multiple
minijets may become observable at the SSC.

 In our model, jet cross sections eventually saturate at the value
   of the total cross section. Thus, we must give some thought to the
value of $Q$ we should use in computing $\g\g$ total cross sections
from the virtual photon distribution
function  \fvdef.  The logarithm in \fvdef\ comes from an integral over
photon transverse momentum.
Ordinarily, to evaluate total cross sections due
to soft processes, one would cut of this integral at a momentum
characteristic of the soft momentum transfer, of order 1 GeV.
To compute the cross section for a hard process, one would run this
integral up to the momentum transfer of the hard process and, therefore,
take $Q = \ps$.  However, when the cross section for a hard process
with momentum transfer $P$
is comparable to the total cross section, photons with transverse
momenta up to this value contribute strongly to the total cross section,
and we must take $Q\sim P$ also to compute the total cross section.
Using \Psparam, we estimated this value as a function of energy.
Thus, in evaluating virtual photon cross sections for jets with
transverse momentum $\ps$, we choose $Q$ in $f_v(x,Q)$ according to
the prescription:
$$ Q  = \max\big[\ps,Q_H(E), 1\ {\rm GeV}
\big], \qquad{\rm where } \ Q_H = (E/10.0)^{0.43}
                          \ ,
                          \eqn\Qprescript$$
and $E$ is the $\g\g$ center of mass energy in the collision.

   We will refer to the model for the  $\g\g$ hadronic cross section
given in \Amaldifit, \Yieldcalc\ or \Psparam,  \meanjet, and
\sigmaprob \ as the
Reference Model (RM).  We feel that this model is the best compromise
available  between simplicity and plausibility in the theoretical
extrapolation of the $\g\g$ hadronic cross section.  We emphasize that
the results of this model related to jet production are
expected to be uncertain to at least a factor of 2.

At some points in the following section, we will compare
the predictions of this model to two additional models which represent
the extreme behaviors possible for this hadronic cross section.  On
the one hand, there is the Constant Cross Section Model (CC), in which
we take
$$     \sigma(\g\g\to \hadrons) =  300\ \nb \ ,
      \eqn\constantcc$$
independent of energy.   On the other hand, there is a model which we
will call the Minijet Dominance Model (MD):
$$     \sigma(\g\g\to \hadrons) =  300\ \nb\  + \  {1\over 2}\YY(\ps),
      \eqn\DGModel$$
with the choice $\ps = 1.6$ GeV.  This is not exactly the model advocated
by Drees and Godbole;  they omit the constant term, and,
   at the end of
ref. \DGLong, they argue that the jet yield estimate should be modified
in a manner similar to what we have described above.
  However, this model captures the
spirit of the explicit calculations  that they have preformed, in a way
that can be easily compared with our reference point.

\chapter{Hadron production rates}

   Having now specified our model completely, we can make use of it
to predict the rate of hadronic $\g\g$ events to be expected at
future colliders.
 In this section, we will present the
results of applying this model to a variety of specific collider
designs, for center of mass energies of 500 GeV and 1 TeV, both
for $\epem$ and for $\g\g$ collisions.  Our set of sample collider
parameters is given in Tables 1 and 2.   Table 1 gives a set of
designs for 0.5 TeV colliders presented at the 1992 Linear Collider
Conference.\Ref\LCHALF{{\sl LC92 Proceedings}, R. Settles, ed.
              ECFA Report 92/46, 1992.}
Table 2 gives a set of designs corresponding to extensions of the
0.5 TeV machines to 1.0 TeV in the center of mass.\Ref\LCONE{The
1.0 TeV designs are taken from the following sources:  DLC: G. Voss
and T. Weiland, \sl SLAC Beam Line, \bf 22, \rm No. 1, 24 (1992);
JLC: K. Yokoya, private communication; NLC, R. Ruth, private
communication; TESLA:  H. Padamsee, \sl TESLA Calculations Program,
\rm Cornell report (unpublished), 1991.}

    Before beginning the analysis of specific designs, we would
like to present some results
 which appear as general scaling laws, independent of the
details of the collider.    This will also give us an opportunity
to compare our reference model (RM) with the minijet dominance model (MD)
and constant cross section (CC) models defined at the end of the
previous section.

   In a  linear $\epem$ collider,  the   rate of hadronic $\g\g$ events
per bunch crossing is  obtained as a convolution of the photon
spectra from bremsstrahlung  and  beamstrahlung.  If we ignore
beamstrahlung and consider the rate from bremsstrahlung alone, our
results will be independent of the detailed collider design  and,
for a fixed design energy, will simply be proportional to the
luminosity per electron/positron bunch crossing.  This is also true
for the full rate of hadronic events in the case where the machine
 is converted to a $\g\g$ collider by backscattering laser
beams, since, in that case,
the energy distribution of backscattered photons is fixed by the
physics of Compton scattering.  As a reference point close to most
current designs, we will assume a design luminosity of
$$   {\cal L} =   10^{34} \cdot \Bigl({\Ecm\over {\rm 1\ TeV}}\Bigr)^2
  \   {\rm cm}^{-2} {\rm sec}^{-1}\ .
   \eqn\designL$$
In typical designs, this luminosity is divided into pulses  which are
produced at a repetetion rate of roughly $f_{\rm rep}\sim 100$/sec.
  In the most recent designs,
which have been inspired by attempts both to raise the design
luminosity and to reduce the Drees-Godbole
background, the electron and positron pulses are divided into trains
of order $n_b\sim 100$ bunches, which we will assume can be
distinguished in time by the detector.
Thus, we take as our
reference value  a luminosity per bunch crossing equal to $10^{-4}$
sec  times \designL, that is,
$$   {\cal L}_1={\cal L}/f_{\rm rep}\cdot n_b
=   10^{-3} \cdot \Bigl({\Ecm\over {\rm 1\ TeV}}\Bigr)^2
  \   {\rm nb}^{-1} .
   \eqn\designLb$$
For any specific machine, the results for bremsstrahlung- or laser
photon-induced hadronic backgrounds can be obtained by scaling the
luminosity per bunch crossing up or down from this value.

  The assumption that the hadrons produced at each bunch crossing
can be distinguished in time is crucial to our analysis and deserves
some further comment.  This assumption is  more or less restrictive
depending on which of the specific collider designs in the Tables is
being considered.
 In designs such as TESLA,
 based on superconducting RF cavities, the bunch spacing is
typically of order 1 $\mu$sec, and there is no problem timing tracks
to much higher accuracy.  However, in the designs based on
conventional cavities, the length of a bunch train cannot be greater
than a  few  hundred nsec, and so the spacing of bunches must be
proportionately smaller.  In the NLC design, for example,
the bunch spacing is only 1.4 nsec.  However, we do not feel that this
is unreasonably small.
The drift chamber of the Mark II
experiment at the SLC could time tracks to a resolution of 1 nsec,
even though it was not optimized for this feature.
Energy clusters in a calorimeter can be given time stamps with 1 nsec
resolution or better by adding
layers of timing detectors, such as scintillation counters,
to the calorimeter.
The time between bunch train crossings is quite long (5.6 milliseconds
and 11.1 milliseconds for the 500 GeV and 1000 GeV NLC designs
respectively) so that one can make use of
timing detectors with large recovery times.

\FIG\eeEnergy{Comparison of
the predictions of three models of the
        $\g\g$ total cross section for the rate of hadronic background
          events in $\ee$ colliders.  Beamstrahlung is ignored, and the
          luminosity per bunch crossing is taken to have the
             canonical dependence \designLb:  (a)  predictions of
               the RM, MD, and CC models (described in the text) for
               the total rate of $\g\g$ events; (b) predictions of
               the RM and MD models for the rate of events with
               observable minijets of transverse energy 5 and 10 GeV.}
    Now we present our estimates of $\g\g$ background rates for
the reference machine defined above, as a function of its energy.
We consider first $\ee$ colliders, ignoring beamstrahlung.  In Fig.
\eeEnergy(a), we plot the  total rate of  $\g\g$ background events,
as a function of the design energy of the machine, assuming the
specific luminosity  \designLb.  The $\g\g$ cross section is integrated
over all solid angle, and down to a $\g\g$ center of mass energy
$E_{\g\g}$ of 5 GeV.  Notice that the MD model predicts a much higher
level of background, while the RM and CC models  are actually quite
close in their predictions.   Under the assumptions of the RM model,
and assuming that the multibunch operation called for in \designLb\ is
indeed feasible, the total rates of hadronic background seem to be
 tolerable
without a need for further  analysis
for $\epem$ colliders of energy
 up to 2 TeV. Unfortunately, the assumption
of ignoring beamstrahlung breaks down well before this point.

 In Fig. \eeEnergy(b), we
show the corresponding predictions for hadronic events with observable
QCD jets, using the MD and RM models with transverse energies above 5
GeV (computed at $\ps = 3.2$ GeV) and 10 GeV (computed at $\ps =8$ GeV).
Again, we integrate over $E_{\g\g} > 5$ GeV.
As the jet transverse momentum increases, the MD and RM models come into
closer agreement.  In addition, the number of events is substantially
smaller, especially at $\ee$ energies of 1 TeV and below.

  At this point in the analysis, it is not clear whether the
true figure of merit for assessing the hadronic backgrounds at
linear colliders is given by the total rate of hadronic events or
only the rate for events containing jets.  In Section 5, we will
report simulation results which indicate that both rates play a
role in determining the hadronic backgrounds.  Events with jets
are  more effective in depositing unwanted background
energy, but events of the minimum bias type can also have some
effect.  As we proceed to discuss specific collider designs, we will
present the total rate of hadronic events and also the rate of
jet events  for $\ps = 3.2$ and 8 GeV.  Taking these three numbers
together, one can obtain a feel for  the general character of the
hadronic background.

\FIG\laserEnergy{Comparison of
the predictions of three models of the
        $\g\g$ total cross section for the rate of hadronic background
          events in $\g\g$ colliders. The conventions are as in Fig.
                \eeEnergy.}
  In Fig. \laserEnergy(a) and (b), we show the results of calculations
  similar    to those of the previous figure
 for $\g\g$ colliders.  Again we assume the luminosity
per bunch \designLb; essentially, we are assuming that high energy
electrons can be converted 1-to-1 to photons.   As Drees and Godbole
pointed out, the results for this case are about an order of magnitude
higher than the bremsstrahlung contribution
for 500 GeV machines, rise faster with energy,
and considerably more model dependent. It is comforting, at
least, that, according to our reference model,
 a 500 GeV $\g\g$ collider
based on most current $\ee$ collider designs should not have a serious
problem with its $\g\g$ background.

   \FIG\ISpec{Spectrum of $\g\g$ hadronic events in the $\g\g$
      center of mass energy,
   $d n/d\log E_{\g\g}$,
          produced in $\ee$ colliders at the
           canonical luminosity \designLb, from brems\-strah\-lung
             photons only.  The three curves represent all $\g\g$
              events, events with 5 GeV minijets, and events with
                 10 GeV minijets. The  two cases are:
                   (a)  500 GeV collider; (b)
               1 TeV collider.}
   \FIG\IGSpec{Spectrum of $\g\g$ hadronic events in the $\g\g$
      center of mass energy,
   $d n/d\log E_{\g\g}$,
       produced in  backscattered-laser
         $\g\g$ colliders at the
           canonical luminosity \designLb.  The three curves are
              as in Fig. \ISpec.  The two cases are:
           (a)  500 GeV collider; (b)
               1 TeV collider.}
  It is interesting not only to know the total number of $\g\g$ hadronic
events but also their distributions in the various kinematic variables.
Of these, the most important is the distribution in the $\g\g$
center of mass energy $E_{\g\g}$, since this quantity determines
the multiplicity of hadrons in the underlying event.  In Figs. \ISpec\
and \IGSpec,
we display the center of mass energy spectrum
$$     E_{\g\g} { d n \over d E_{\g\g}}
    \eqn\specdef$$
per bunch crossing for the canonical machine design described
above, for $\Ecm$ = 500 GeV and 1 TeV $\ee$ and $\g\g$
colliders.   These calculations asssume the Reference Model.
Notice that the largest number of background events in $\ee$ colliders
involve relatively low-energy $\g\g$ scattering processes.  On the
other hand, in a $\g\g$ collider, the luminosity spectrum of the
background, like the spectrum of signal processes, peaks at the
highest available energy.

   \FIG\FCSpec{Spectrum of $\g\g$ hadronic events, for two
       representative designs for 500 GeV $\ee$  colliders,
          following the parameters given in Table 1.
        The three curves are as in Fig. \ISpec.  The dotted curves
           are the corresponding results for bremsstrahlung only. The
          two cases considered are  (a) JLC; (b) TESLA.}
   For $\epem$ colliders, beamstrahlung is an important source of
photons. Unfortunately, the results both for the number and the
spectrum of beamstrahlung photons depend on the details of the
machine design, and, in particular, on the number of particles per bunch
and the bunch geometry. The disruption effect during beam-beam collision
further complicates the situation, as discussed in Section 2.
  Thus, to assess the $\g\g$ backgrounds due
to beamstrahlung, we must work with specific parameter sets for
proposed colliders.  In Table 1, we list six proposed parameter
sets for 500 GeV colliders.\Ref\noVLEPP{For the case of VLEPP,
     the prescription of Section 2 fails due to the large value
of    $\sigma_z/\beta^*_y$.  The correct remedy (V. Balakin,
private communication) is to set $H_{D_y}=1$ and proceed from
eq.  \sigmaeff.}
  For each of these, we have computed
the number of $\g\g$ hadronic collisions per bunch crossing.
In the table, we quote the values of $N_{\rm had}$,
the total number of hadronic events,
$N_{\rm jet5}$, the number of hadronic events with 5 GeV minijets
(computed at $\ps=3.2$ GeV), and $N_{\rm jet10}$,
 the number of events with
10 GeV minijets (computed at $\ps =8$ GeV).  Each of these numbers
is integrated over the range $E_{\g\g}> 5$  GeV.
   We also quote the
values of these parameters for the case in which the electron beams
are converted to  backscattered photon beams, assuming no loss of the
nominal luminosity.\Ref\whichLumi{For
$\ee$ colliders, we estimate event rates using
the enhanced
luminosity per bunch
$\bar{\cal  L}_1$.  For $\g\g$ colliders, we use the unenhanced
luminosity ${\cal L}_1$.}
In Fig. \FCSpec,
We show the distribution of $E_{\g\g}$ for two representative cases,
JLC and TESLA.   The corresponding
spectra for the photon colliders can be obtained by scaling from
Fig. \IGSpec(a), as we have remarked above.

We see from Table~1 that  disruption effects have two
    major impacts of disruption on beamstrahlung and the $\g\g$
backgrounds.   First, disruption  hardens the beamstrahlung
spectrum and increases its radiation rate. In addition,
 disruption  enhances
the luminosity per bunch crossing. In machine designs such as CLIC, DLC,
  and TESLA, for which the
beams are not extremely flat, the horizontal disruption $D_x$ can be
quite large.
 This leads to an effective luminosity and concomitant
 beamstrahlung  substantially
different from the nominal designed values. Care must be taken to
include these effects when evaluating beamstrahlung and the backgrounds.

In estimating the size of the beamstrahlung-induced  backgrounds,
one should pay special attention to the parameter
$n_{\g}$, the average number of beamstrahlung photons
radiated per electron. Since in the collider energy range of our interest
the hadron total cross section is reasonably constant in the $\g\g$
center-of-mass energy, the total hadronic event rate $N_{\rm had}$
scales roughly as the square of  $n_{\g}$ when ${\bar{\cal L}}_1$ is
fixed. This is the source of the variation by almost two orders of
magnitude among the first five
machines of Table~1 in the total rate of hadronic
events per bunch crossing.
Since the  minijet production comes dominantly from high energy
photons, the jet cross sections are less sensitive to $n_\g$.
The sixth machine,
  VLEPP, has a very different design philosophy from the first five
in having
only one bunch in a pulse train. This results in a luminosity per bunch
crossing which is about 100 times larger than all other machines. When
convoluted with a large number of beamstrahlung photons (mainly due to
a large $\sz$), VLEPP tends to produce the most hadronic event rates
among all the machines.

  Since there are no beamstrahlung  and disruption
effects involved in the $\g\g$ collision,
 the hadronic backgrounds in the $\g\g$ mode
are very comparable among
the first five machines listed in Table~1.

   \FIG\FCTspec{Spectrum of $\g\g$ hadronic events, for two
       representative designs for 1 TeV $\ee$  colliders,
          following the parameters given in Table 2.
        The three curves are as in Fig. \ISpec.  The dotted curves
           are the corresponding results for bremsstrahlung only. The
        figures correspond to (a) JLC; (b) TESLA.}
  In Table 2, we present some representative designs for 1 TeV colliders,
 and our estimate of the $\g\g$ background rates both in $\ee$ and
$\g\g$ collider modes.  For the $\ee$ collider designs, the
 $E_{\g\g}$ spectra for the four machines
are shown in Fig. \FCTspec. Again in the case of DLC and TESLA designs,
the relatively larger disruption effects and $n_{\g}$
 lead to a much higher rate of hadronic events.
Although the $\g\g$
reactions are mainly soft, one finds more than one jet-like underlying
event per bunch crossing in these two cases.
In the 1 TeV $\g\g$ collision mode, it is true for all four machines
that typical
events have underlying hadronic events with QCD minijets. It is
interesting to note that, with the differences from beamstrahlung
removed in the $\g\g$  mode,
the hadronic and minijet event rates for
 DLC and TESLA are comparable to those for
 JLC and NLC.

\chapter{Simulation of hadronic backgrounds}

   Now that we have computed the fraction of $\ee$ or $\g\g$ events
at a linear collider which have underlying hadronic activity, we should
still ask how these underlying hadronic events affect the analysis
of high-energy event on which they may be superposed.  There are
good reasons to expect that the answer to this question should
further diminish the importance of the Drees-Godbole background
processes.  The $\g\g$ collisions whose rates we computed in the
previous section typically occur between photons of  unequal
energy, leading to a highly boosted final hadronic system.
Even if this system contains minijets, it will include relatively few
high transverse momentum particles.  Thus, most of the final hadrons
will disappear down the beam pipe in the forward or backward
direction.  Unfortunately, we did not find a simple way to estimate
the expectation for the resulting energy distributions, except by
direct simulation.  In this section, we will describe the results of
a simulation of the background hadronic energy deposition based on
our Reference Model.

  According to Fig. \Njets, the qualitative form of hadronic background
events will be different for the different sets of colliders we have
considered.  For the 500 GeV electron colliders, not only are there
relatively few hadronic events per bunch crossing, but also those
hadronic events are typically of minimum bias type without minijets.
On the other hand, the background events at a 1 TeV $\g\g$ collider
typically contain several minijet pairs, depending on the transverse
momentum criterion for a distinct minijet.  We will present results for
both of these cases.

   To simulate the hadronic events, we first generate minijet
pairs according to the Reference Model, using a Poisson distribution
with mean given by   \meanjet.  The  energy and angle distributions
of the minijets are computed according to \Yieldcalc, in particular,
using the gluon and quark distributions in the photon given by
Drees and Grassie.\refmark\DGrass   This calculation requires a cutoff
$\ps$
on the transverse momentum transfer in the parton-parton collision;
we have taken this cutoff to be either 3.2 or 8 GeV.  As we explained
in Section 3, the first
of these values would assign the production of clusters of 5 GeV
transverse momentum to incoherent production of minijets, while the
second value would make this assignment only for clusters of 10 GeV
transverse momentum.
The partons are fragmented
into jets using the Lund Monte Carlo, version 6.3 .\Ref\Lund{
T.~Sj\"ostrand, \sl Comput. Phys. Comm. \bf 39, \rm 347 (1986);
M.~Bengtsson and T.~Sj\"ostrand, \sl Nucl. Phys. \bf B289, \rm
810 (1987).}
Following the formation of all minijet pairs, the remaining energy
is converted to hadrons using the minimum bias event generator of the
ISAJET Monte Carlo.\Ref\IsaJet{
F. E. Paige and S. D. Protopopescu, in
{\sl  Proceedings of the 1986 Summer Study on the
Physics of the Superconducting Supercollider},
R. Donaldson and J. Marx, eds. (Fermilab, Batavia,IL,1986).}
We modify this generator
only in replacing the leading baryons by $\rho^0$'s.   We restrict
our analysis to events with hadronic invariant mass at least 5 GeV.

   To simulate the response of a detector to these events, we have
used a model based on the features and resolution of the SLD
detector.\Ref\Sld{The SLD Design Report, SLAC Report 273, 1984.}
As an important modification from the design of the SLD, we have
assumed that the detector is blind to particles passing within
10$^\circ$ of the beamline ($|\cos\theta| > 0.985$). Planned detectors
for future linear colliders include masking in this region to
control the effects of electron-positron pairs produced in the
collisions of electron and positron bunches, and simulations
of physics signals at
 linear colliders incorporate  this angular constraint.

   From this model, we compute $\Edep$, the total charged and neutral
energy deposited in the detector by a single hadronic background event.
We also compute a number of subsidiary quanties:  To assess the effect
of a stronger angular restriction, we have examined the quantity
 $\Edep(0.9)$,  the
energy deposited in the angular region $|\cos\theta| < 0.9$.
Most physics analyses for future linear colliders are insensitive
to this restriction.  Since missing transverse momentum signatures are
important for some physics processes, we have computed $\Ptmiss$,
the missing
transverse momentum  observed by the detector for the
hadronic event.  Finally, we have recomputed the observed missing
transverse momentum using the stronger angular restriction, to define
$\Ptmiss(0.9)$.

    The results of this analysis are shown in Table 3 for three
    cases which represent the range of possibilities:
  first
 the background hadronic events at a 500 GeV
$\ee$ collider, second, the background hadronic events at a 1 TeV
$\ee$ collider (in both cases,
assuming the NLC design) and, finally,  the hadronic events
from monochromatic $\g\g$ collisions at 1 TeV in the center of
mass. Note that the last case involves harder $\g\g$ collisions than
one would find from the photon spectrum \backscatter.
 In the first two cases, it is relatively rare that a hadronic
event will contain a minijet pair, and so there is little difference
between the results with $\ps = 3.2$ GeV and $\ps = 8$ GeV.  In the
third  case, however,
a typical hadronic event contains a 5 GeV minijet pair.
Thus, there is a considerable difference between the results for the
two parameter choices, and this difference mainly reflects the explicit
inclusion of the minijets in the former case.  We have presented the
results from both choices for comparison.

 \FIG\EEcollEspec{Distribution of deposited energy for hadronic
         background events for a 500 GeV $\ee$ collider with
         a detector angular coverage of $|\cos\theta|<0.985$ (solid)
         and of $|\cos\theta|<0.9$ (dashed).
          The deposited energy is the sum of charged and neutral
           energy recorded by a detector, which is modeled as described
          in the text.  The number of events corresponds to an
            integrated luminosity of 0.5 pb$^{-1}$.}
 \FIG\GGcollEspec{Distribution of deposited energy for hadronic
   events in 1 TeV monochromatic $\g\g$ collisions,  (a) computed
      with $\ps = 3.2$ GeV, (b) computed with $\ps = 8$ GeV.
           The number of events corresponds to an
            integrated luminosity of 10 nb$^{-1}$.}

 For the background events at $\ee$ colliders, we were surprised by
the  small values that the simulation produces, both for the
energy deposition and for the missing transverse momentum.
In Fig. \EEcollEspec, we display the distributions of $\Edep$
and $\Edep(0.9)$
computed for the
500 GeV NLC collider.  The distributions fall off exponentially, with
mean energy depositions of 7.9 GeV and 3.3 GeV for $\Edep$ and
$\Edep(0.9)$ respectively.
   The event numbers in the
histogram of Fig. \EEcollEspec\ correspond to an integrated
luminosity of 0.5 pb$^{-1}$.  This yields 40,000 hadronic background
events, of which none has $\Edep(0.9)$ greater than 50 GeV.  If we
include in the simulation hadronic events with invariant mass
down to 1.6 GeV, this adds another 33,000 hadronic events, of which
virtually all have $\Edep$ less than 8 GeV and
$\Edep(0.9)$ less than 4 GeV.   The missing transverse
momentum in our sample of hadronic events is typically
less than 1 GeV.  Since the physics processes for which this is a
signature typically have missing transverse momentum of order
$m_W$, this small uncertainty is quite unimportant. The qualitative
features of the hadronic background events, including the
exponentially falling distribution in deposited energy, are
the same for the 1 TeV $\ee$ collider.

  For 1 TeV $\g\g$ collisions, the hadronic background  events have a
more serious effect, and one which depends more strongly on the
model used to generate these events.  In Fig. \GGcollEspec, we
show the distribution in deposited energy for these events corresponding
to 10 nb$^{-1}$ of integrated luminosity.  In Fig. \GGcollEspec(a),
we generate hadronic events using $\ps = 3.2$ GeV.  Then, typical
events contain at least one minijet pair.  In Fig. \GGcollEspec(b),
we assume $\ps = 8$ GeV, corresponding to the more restrictive
hypothesis that minijets are not produced incoherently from the
minimum-bias hadronic production mechanisms until they acquire a more
substantial transverse momentum.  We put forward both calculations to
illustrate the possibilities.  At a working $\g\g$ collider, it will
of course be straightforward to measure this background and model it
accurately.  In either model, the background events deposit
significant amounts of energy, but mostly in the extreme forward
and backward directions.  They add relatively little uncertainty to
the determination of missing transverse momentum.

  We argued in the previous section that, at $\ee$ linear colliders
up to 1 TeV, only a small fraction of the $\ee$ annihilation
events should have underlying background hadronic events generated
by $\g\g$ collisions.  Now it seems that, when such a background event
does appear, it makes only a minor modification of the event pattern
of the $\ee$ annihilation.  For $\g\g$ colliders, especially at 1 TeV,
the background problem is more serious.  Though the  modification of
 the energy deposition in typical events is small, it is signficant and
should be  taken into account in physics simulations.

\chapter{Conclusions}

   Drees and Godbole have called attention to the large rate of
   photon-photon collisions to be expected at future linear colliders
   and have suggested that the presence of underlying $\g\g$ events
   might be a serious source of background.  To evaluate this claim,
    there are two issues that must be understood.

  First, one must carefully evaluate the
   expected rate of soft and jet-like $\g\g$ events to be expected
   for given collider parameters.  In this paper, we have presented
      what we feel is a useful solution to this problem.
   We have presented a
 physically correct picture of
    the hard and soft components of the $\g\g$ total cross section,
   and we have provided a set of formulae which allows this picture
    to be applied straightforwardly to compute the $\g\g$ rate
     for any collider design.

      Second, one must ask whether such
underlying hadronic events actually affect the experiments to
be carried out at the next generation linear colliders.
It is possible that any underlying
event will compromise some aspect of linear collider physics.  We have
presented simulation results which indicate that the effect of this
background will be minor at 500 GeV $\ee$ colliders, but that it will
be more significant at higher energies, especially in $\g\g$ collision
mode.   Even in 1 TeV $\g\g$ collisions, the Drees-Godbole background
remains a relatively small perturbation of a high energy reaction.
In any event,
we have given prescriptions which allow the Drees-Godbole
background to be correctly included in simulations of physics
processes to assess its effects directly.

As a final note,
we should   point out that there are strategies for reducing the
Drees-Godbole backgrounds   in $\ee$ colliders
 by readjusting the beam
parameters. For example, by increasing the collision rate
while lowering the bunch population, and by colliding extremely
flat beams,
one may decrease the $\g\g$ reaction rates
while retaining the total luminosity. We hope that the
case studies and approximation schemes presented in this paper
will be useful in further optimizing the designs for linear colliders.

\ACK

We are deeply grateful to James Bjorken for helping us to understand
the basic physical picture which is used in this paper.  We also
thank Karl Berkelman, Manuel Drees,  Rohini Godbole, John Storrow,
and Valery Telnov for enlightening  discussions.

\endpage
\input tables
\vsize= 9.0 in
\hsize= 6.0 in
\baselineskip= 14 pt
\centerline{Table 1. Parameters and Hadronic Backgronds
            for 0.5 TeV  Linear Colliders}
\bigskip
\begintable
 Linear Colliders    |CLIC | DLC  | JLC  | NLC | TESLA | VLEPP   \crthick
 ${\cal L}[10^{33}\lunit]$\hfill |2.7|2.4| 6.8 | 6.0 | 2.6 |  12    \nr
 $f_{\rm rep}$[Hz]\hfill  | 1700|  50  | 150  | 180 |  10  |  300     \nr
 $n_b$ \hfill         |  4  | 172  |  90  |  90 | 800  |    1     \nr
 ${\cal L}_1[10^{-3}{\rm nb}^{-1}$]\hfill|0.40| 0.27
  |0.50 |0.37 |0.33 |  40    \nr
 $N$[$10^{10}$]\hfill| 0.6 |  2.1  | 0.7 | 0.65 | 5.15|    20    \nr
 $\sx/\sy$[nm] \hfill |90/8 | 400/32|260/3|300/3|640/100|2000/4    \nr
 $\sz[\mu m]$\hfill  | 170 | 500 | 80   | 100| 1000 | 750     \nr
 $\beta_x^*/\beta_y^*$[mm] \hfill| 2.2/0.16 | 16/1| 10/0.1|10/0.1|
                     10/5   | 100/0.1 \crthick
 $\Dx/\Dy$ \hfill |1.3/15|0.70/8.8|0.09/8.2|0.08/8.2|1.25/8.0|0.43/---\nr
 $A_x/A_y$ \hfill|0.08/1.06|0.03/0.5|0.008/0.8|
                   0.01/1.0|0.1/0.2|0.008/---\nr
 $\bar{\sigma}_x/\bar{\sigma}_y$[nm]\hfill |40/5.5|246/19|259/2.0|
                                     300/2.2|304/50 |1587/4\nr
 $H_{_D}$\hfill     |  3.3| 2.8|  1.5 | 1.4| 4.2  | 1.3 \nr
 $\bar{\cal L}[10^{33}\lunit]$\hfill |8.80|6.67  |
                                         10.1 | 8.22|11.1 |15.1\nr
 $\bar{\cal L}_1[10^{-3}{\rm nb}^{-1}$]\hfill|1.30| 0.76
  |0.74 |0.51 |1.39 |  50.2  \crthick
 $\Y_0$ \hfill         | 0.16| 0.043| 0.15| 0.095| 0.031 | 0.059\nr
 $\Y$ \hfill      |0.35 |0.071 | 0.15  |0.096 |0.065  |0.074 \nr
 $\delta_B$ \hfill   | 0.36| 0.08| 0.05| 0.03 | 0.14| 0.14 \nr
 $n_{\g}$ \hfill   | 4.6 | 3.1      |1.0  |  0.84|5.8  |5.1 \crthick
 $\epem$ Mode      |     |     |       |       |     |     \cr
 $N_{\rm had}$\hfill |1.37| 0.32 | 0.07 | 0.04 |1.57 | 45.3    \nr
 $N_{\rm jet5}[10^{-2}]$\hfill |5.80|0.44  |0.22   |
          0.10  | 1.62 | 56.2   \nr
 $N_{\rm jet10}[10^{-4}]$\hfill |16.4 |1.16 |0.69|
          0.31 | 3.90|139     \crthick
 $\g\g$ Mode      |     |     |       |       |     |     \cr
 $N_{\rm had}$\hfill |0.15| 0.10 | 0.19 | 0.14 |0.13 | 15.2    \nr
 $N_{\rm jet5}[10^{-2}]$\hfill |6.90|4.72  |8.61   |
          6.43  | 5.68 |685   \nr
 $N_{\rm jet10}[10^{-4}]$\hfill | 32.4| 22.3|40.7|
           30.4|26.9 |3240   \endtable
\endpage
\centerline{Table 2. Parameters and Hadronic Backgrounds
            for 1.0 TeV  Linear Colliders}
\bigskip
\begintable
 Linear Colliders            | DLC  | JLC | NLC  | TESLA    \crthick
 ${\cal L}[10^{33}\lunit]$\hfill |  2.5 | 8.8 | 12.8 |  10.6  \nr
 $f_{\rm rep}$[Hz]\hfill     |  50  | 150 |  90  |  10    \nr
 $n_b$ \hfill                |  50  |  20 |  90  | 800    \nr
 ${\cal L}_1[10^{-3}{\rm nb}^{-1}$]\hfill|0.99| 2.17
                                          | 1.58  | 1.31  \nr
 $N$[$10^{10}$]\hfill        |  2.8 | 1.8 | 1.3  | 5.8   \nr
 $\sx/\sy$[nm] \hfill       | 223/28.3|372/3.2|425/2 |404/50.5  \nr
 $\sz[\mu m]$\hfill          |  500 | 113 | 100  | 1100    \nr
 $\beta_x^*/\beta_y^*$[mm] \hfill| 5/0.8|24.6/0.12|40/0.1|8/2.5 \crthick
 $\Dx/\Dy$ \hfill      |1.40/11.0 | 0.08/9.7| 0.04/8.5 | 1.95/15.6  \nr
 $A_x/A_y$ \hfill      |0.1/0.625| 0.005/0.9| 0.0025/1.0| 0.14/0.44  \nr
 $\bar{\sigma}_x/\bar{\sigma}_y$[nm]\hfill | 100/17.1 | 372/2.2
                                  | 425/1.5 | 172/27.0 \nr
 $H_{_D}$\hfill                | 3.7  | 1.5   | 1.4 | 4.4  \nr
 $\bar{\cal L}[10^{33}\lunit]$\hfill | 9.2  | 12.8 | 17.5 | 46.6 \nr
 $\bar{\cal L}_1[10^{-3}{\rm nb}^{-1}$]\hfill|3.70| 3.10
                                          | 2.18  | 5.86  \crthick
 $\Y_0$ \hfill                 | 0.20  | 0.38  | 0.27 | 0.10 \nr
 $\Y$ \hfill                   | 0.42  | 0.38  | 0.27 | 0.24 \nr
 $\delta_B$ \hfill             | 0.53  | 0.14  | 0.07 | 0.50 \nr
 $n_{\g}$ \hfill               | 8.1   | 1.7   | 1.1  | 10.4  \crthick
 $\epem$ Mode                  |       |       |      |     \cr
 $N_{\rm had}$\hfill           | 15.3  | 0.83  | 0.34 | 40.1 \nr
 $N_{\rm jet5}$\hfill          | 1.53  | 0.09  | 0.03 | 2.65 \nr
 $N_{\rm jet10}[10^{-2}]$\hfill| 5.53  | 0.37  | 0.12 | 8.54 \crthick
 $\g\g$ Mode                   |       |       |      |     \cr
 $N_{\rm had}$\hfill           | 0.42  | 0.93  | 0.68 | 0.56 \nr
 $N_{\rm jet5}$\hfill          | 0.31  | 0.68  | 0.50 | 0.41 \nr
 $N_{\rm jet10}[10^{-2}]$\hfill| 2.50  | 5.61  | 4.10 | 3.40 \endtable
 \endpage
\centerline{Table 3. Simulation Results on Hadronic Backgrounds}
\bigskip
\begintable
 Linear Colliders            | NLC (500 GeV) | NLC (1 TeV) |
    $\g\g$ (1 TeV) | $\g\g$ (1 TeV)  \nr
 $\ps$  |   3.2   | 3.2  |  3.2  | 8.0 \crthick
 $\Edep$ (GeV)        | 8.0   | 11.0   |  64. | 33.  \nr
 \quad st. dev.       | 7.1   | 11.6   |  31. | 25.  \cr
 $\Edep(0.9)$ (GeV)   | 3.3   |  4.4   |  25. | 12.  \nr
 \quad st. dev.       | 3.3   |  5.0   |  13. | 11.  \crthick
 $\Ptmiss$ (GeV)      | 0.6   | 0.9    |  4.7 | 2.4  \nr
 \quad st. dev.       | 0.5   |  0.9   |  3.2 | 2.9  \cr
 $\Ptmiss(0.9)$ (GeV) | 0.7   |  0.9   |  5.0 | 2.3  \nr
 \quad st. dev.       | 0.6   |  1.0   |  3.6 | 3.1  \endtable
\endpage
\refout
\endpage
\figout
\endpage
\end